\documentclass[journal]{IEEEtran}

\usepackage{graphicx}
\usepackage{subcaption}
\usepackage{url}

\usepackage{amsmath, amssymb}
\usepackage[ruled,vlined]{algorithm2e}
\usepackage{algorithmic}

\usepackage{subcaption}
\usepackage{url}
\usepackage[ruled]{algorithm2e}
\usepackage{algorithmic}
\usepackage{multicol}
\usepackage{multirow}
\usepackage{caption}
\usepackage[dvipsnames]{xcolor}
\usepackage{collcell}
\usepackage{wrapfig}  
\makeatletter
\def\hlinewd#1{%
  \noalign{\ifnum0=`}\fi\hrule \@height #1 \futurelet
   \reserved@a\@xhline}
\makeatother

\makeatletter
\newcommand{\removelatexerror}{\let\@latex@error\@gobble}

\begin{document}

\title{RecSys-DAN: Discriminative Adversarial Networks for Cross-Domain Recommender Systems}

\author{Cheng Wang,~\IEEEmembership{Member ~IEEE}, Mathias Niepert, ~Hui Li$^{*}$ 
% <-this % stops a space 
\thanks{$^{*}$Corresponding author.} 
\thanks{Cheng Wang is with the NEC Laboratories Europe, Heidelberg, Germany, e-mail: cheng.wang@neclab.eu.}
\thanks{Mathias Niepert is with the NEC Laboratories Europe, Heidelberg, Germany, e-mail: mathias.niepert@neclab.eu.} 
\thanks{Hui Li is with the Fujian Key Laboratory of Sensing and Computing for Smart City, School of Information Science and Engineering,
Xiamen University, Xiamen, Fujian, P. R. China. e-mail: hui@xmu.edu.cn.}% <-this % stops a space 
% \thanks{Manuscript received April 19, 2005; revised August 26, 2015.}
}

% \markboth{Journal of \LaTeX\ Class Files,~Vol.~14, No.~8, August~2015}%
% {Shell \MakeLowercase{\textit{et al.}}: Bare Demo of IEEEtran.cls for IEEE Journals}

\maketitle

\begin{abstract}
Data sparsity and data imbalance are practical and challenging issues in cross-domain recommender systems.  This paper addresses those problems by leveraging the concepts which derive from representation learning, adversarial learning and transfer learning (particularly, domain adaptation). 
Although various transfer learning methods have shown promising performance in this context, our proposed novel method RecSys-DAN focuses on alleviating the cross-domain and within-domain data sparsity and data imbalance and learns transferable latent representations for users, items and their interactions. Different from existing approaches, the proposed method transfers the latent representations from a source domain to a target domain in an adversarial way. The mapping functions in the target domain are learned by playing a min-max game with an adversarial loss, aiming to generate domain indistinguishable representations for a discriminator. Four neural architectural instances of ResSys-DAN are proposed and explored. Empirical results on real-world Amazon data show that, even without using labeled data (i.e., ratings) in the target domain, RecSys-DAN achieves competitive performance as compared to the state-of-the-art supervised methods. More importantly, RecSys-DAN is highly flexible to both unimodal and multimodal scenarios, and thus it is more robust to the cold-start recommendation which is difficult for previous methods.
\end{abstract}


\IEEEpeerreviewmaketitle

\section{Introduction}
\label{sec:intro}

\IEEEPARstart{R}{ecommender} systems (RS) generate predictions based on the
customers' preferences and purchasing histories. Collaborative filtering
(CF) and content-based filtering
(CBF) are popular techniques used in
such systems~\cite{koren2009matrix}.  CF-based methods generate recommendations by computing latent
representations of users and products with matrix factorization (MF) methods~\cite{LiCYM17}.  Although CF-based
approaches perform well in several application domains, they
are based solely on the \emph{sparse} user-item rating matrix and, therefore,
suffer from the so-called \textit{cold-start} problem \cite{schein2002methods}. For new
users without a rating history and newly added products with few or no ratings
(i.e., sparse historical data), the systems fail to generate high-quality
personalized recommendations.

\begin{figure}[!t]
  \begin{center}
    \includegraphics[width=0.65\columnwidth]{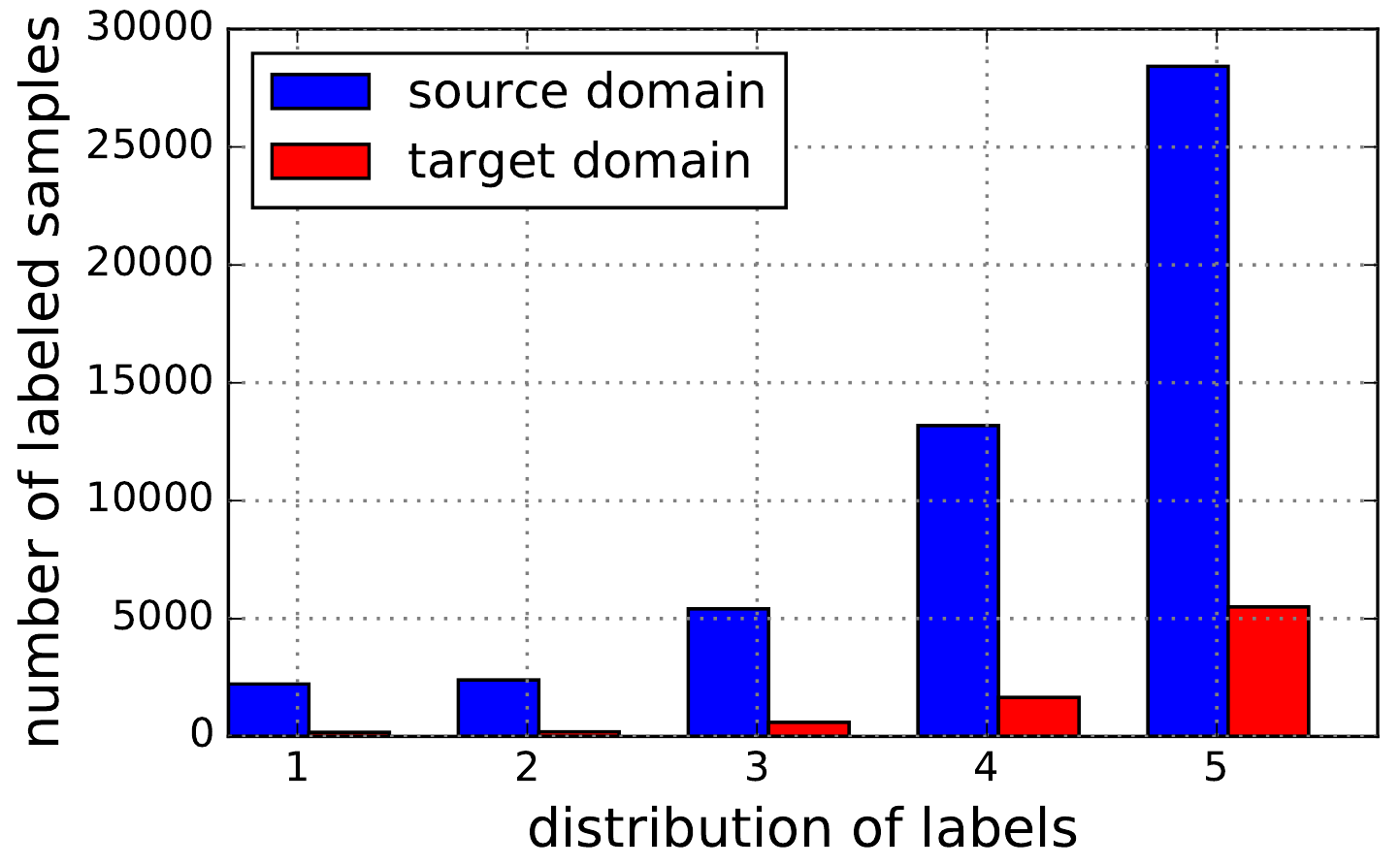}
  \end{center}
\vspace{-2mm}
\caption{\label{fig:rating_dis} The illustration of cross-domain imbalance and within domain imbalance problems in cross-domain recommendation problem. The \textit{source domain} represents product domain ``digital music" and \textit{target domain} stands for product domain ``music instrument'' in Amazon dataset (see Section~\ref{sec:data} for the detailed explanation of the dataset).}
\vspace{-5mm}
\end{figure}

Alternatively, CBF approaches
leverage auxiliary information such as product
descriptions~\cite{WangNL18}, 
locations~\cite{Lu17HBGG} and social network~\cite{LiWTM15} to generate
recommendations. These methods are in principle more robust to cold-start
problem as they can utilize different modalities.  However, a pure CBF
approach will face difficulties in learning sharable and transferable
information of users and items across different product domains (e.g., ``book"
or ``movie")~\cite{pan2010survey}. % through some form of transfer learning.
A typical example of this scenario is cross-domain
recommendation.  Large online retailers such as
Amazon and eBay often obtain user-item preferences from multiple domains so
that the quality of recommendation could be improved by transferring knowledge
acquired in a source domain to a target domain. The source-target data domain
pairs in cross-domain recommendation are typically \emph{imbalanced} in two
aspects: \textit{cross-domain imbalance} and \textit{within-domain imbalance}. The former means
that the numbers of users, items or labels in two domains are imbalanced (as shown in Tab.~\ref{tab:dataset}),
The latter refers to the problem that the distribution of categorical
labels (i.e., rating scores) within one domain is imbalanced.
Fig.~\ref{fig:rating_dis} presents the imbalanced scenarios in 5-score based
cross-domain recommendation. In this example, both cross-domain imbalance and
within-domain imbalance exist.

Alleviating the aforementioned data sparsity and data imbalance problems is a non-trivial issue for the cross-domain recommendation. However,  existing CF-based and CBF-based approaches may fail to handle the problems when data becomes more and more sparse. One possible solution is to shift the learning schema from supervised to semi-supervised with limited labeled data. When it comes to a target domain in which the labeled data are completely unavailable, the only way to make a recommendation is transfer learning, particularly domain adaptation, by leveraging the knowledge from other domains.

To address the limitation of existing methods, in this paper,  we propose a method called \emph{Discriminative Adversarial Networks for Cross-Domain Recommendation (RecSys-DAN)} 
to learn the transferable latent representations of users, items and user-item pairs across different product domains. RecSys-DAN is rooted in the recent success of imbalanced learning~\cite{he2008adasyn, he2009learning, he2013imbalanced, ming2015unsupervised, tsai2016domain}, transfer learning~\cite{li2009transfer} and adversarial learning~\cite{goodfellow2014generative}, 
It adopts unsupervised adversarial loss function in combination with a discriminative
objective. 

\begin{figure}
  \begin{center}
    \includegraphics[width=0.375\textwidth]{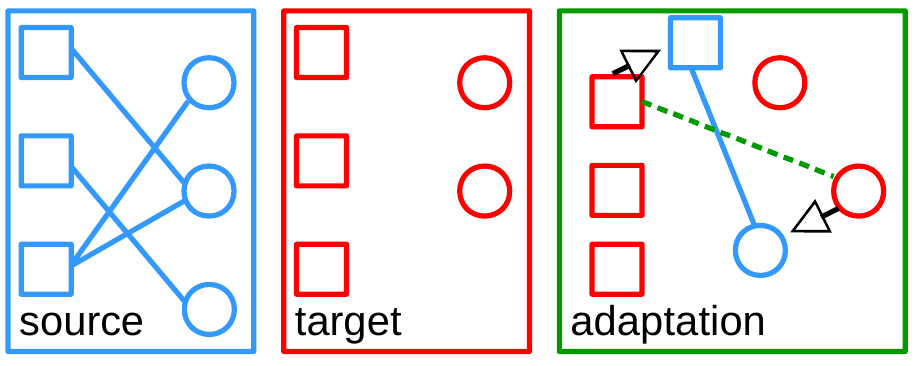}
  \end{center}
\caption{\label{fig:adaptation}Unsupervised adversarial adaptation for cross-domain recommendation. Each square presents a user and circle presents an item, the links between users and items present the preference information (rating) that users express on items. The rating scores are not available in the target domain. The dash links are generated by our proposed method with adversarial adaptation.}
\vspace{-4mm}
\end{figure}

A related research field to RecSys-DAN is domain adaptation~\cite{mansour2009domain}. Although domain adaptation has shown the capability to mitigate the
rating sparsity problem, we argue that adversarial domain adaptation~\cite{ganin2015unsupervised} for recommender systems has two distinct advantages. 
First, with unsupervised adversarial domain adaptation, we can learn a recommendation model when labels in the target domain
are entirely not available, the typical domain adaption usually week or even not work in this case~\cite{man2017cross,liu2018transferable,yu2018svms}. Second, we can observe the performance improvements that brought from adversarial domain adaptation as compared to traditional domain adaption, and we reported the evidence in Tab.~\ref{tab:modality}.
 Moreover, RecSys-DAN incorporates not only rating information but
also additional user and item features such as product images and review texts.
Fig.~\ref{fig:adaptation} demonstrates how RecSys-DAN aligns objects with
different types  and their existing preference relationships in order to
predict new preference relationships in the target domain.

RecSys-DAN targets at the cold-start scenarios where no or only very few user-item preferences are available in the target domain. Existing supervised
methods~\cite{lee2001algorithms, koren2008factorization,
mcauley2013hidden, iwata2015cross,  li2014matching,
liu2015non, he2017neural, zheng2017joint} fail in this setting. We
evaluate RecSys-DAN on real-world datasets and explore various scenarios where
the information in the source and target domains are in the form of uni-modality or multi-modality.  The experimental results show that RecSys-DAN
achieves competitive performance compared to a variety of state-of-the-art
supervised methods which have access to ratings in the target domain.

In summary, RecSys-DAN makes the following contributions:
\begin{itemize}
\item RecSys-DAN is the first neural framework adopting an adversarial loss for the cold-start problem that caused by data sparsity and imbalance in cross-domain recommender systems. It learns domain indistinguishable representations of different types of objects (users and items) and their interactions. 
\item RecSys-DAN is a highly flexible framework,  which incorporates data in various modalities such as numerical,  image and text.  
\item RecSys-DAN addresses the cross-domain data imbalance issue as well as imbalanced preferences in recommender systems by using representation learning and adversarial learning.
\item RecSys-DAN achieves very competitive performance to the state-of-the-art supervised methods on real-world datasets where the target labels are completely not available. 
\end{itemize}

The rest of this paper is organized as follows: Section~\ref{sec:related}
provides background and discusses related work. We present the motivation and
problem statement in Section~\ref{sec:motivation}. The details of our proposed
approach, RecSys-DAN, are illustrated in Section~\ref{sec:method}. Experiments
on real datasets that demonstrate the practicality and effectiveness of
RecSys-DAN are presented in Section~\ref{sec:exp}. Section~\ref{sec:con}
concludes our work.

\section{Related Work} 
\label{sec:related}
This work is related to
four lines of work: cross-domain recommendation, imbalanced learning,
adversarial learning and domain adaptation. 
%In this section, we will elaborate on the related work of these four fields.

\subsection{Cross-domain recommendation}
% Although variety kinds of methods have been proposed to facilitate personalized recommendation, e.g., collaborative filtering (CF)~\cite{koren2009matrix}, content-based filtering (CBF)~\cite{pazzani2007content,lops2011content}, fuzzy approaches~\cite{yera2017fuzzy}, CF and CBF have been proven as effective approaches in the family of recommender systems, they have their own limitations, i.e., data-sparsity for CF and over-specialization for CBF~\cite{lu2015recommender}. 

Cross-domain recommendation (CDR) offers
recommendations in a target domain by exploiting knowledge from source
domains. To some extent, CDR can overcome the
limitations of traditional recommendation approaches. It has been viewed as a
potential solution to mitigate the cold-start and sparsity problem in
recommender systems. Some methods have been proposed~\cite{tang2012cross,
iwata2015cross} along this line. EMCDR~\cite{man2017cross} is proposed to
learn a mapping function across domains. TCB~\cite{liu2018transferable} learns
transferable contextual bandit policy for CDR. 
% CoNet~\cite{hu2018conet}
% connects domains by cross-mappings. 
Sheng et al.~\cite{gao2013cross} propose
ONMTF, which is a non-negative matrix tri-factorization based method. Xu et
al.~\cite{yu2018svms} recently propose a two-side cross-domain model
(CTSIF\_SVMs) which assumes that there are some objects (users and/or items)
which can be shared in the user-side domain and item-side domain. Different
to these methods, RecSys-DAN considers that target domain has completely
unlabeled data (i.e., no ratings). Existing methods will encounter
difficulties in learning effective models for such a scenario.

\subsection{Imbalanced Learning}

Recently, Imbalanced learning~\cite{he2008adasyn, he2009learning,
he2013imbalanced, xue2015does} has been adapted to cross-domain
data~\cite{ming2015unsupervised, tsai2016domain}. Xue et.
al~\cite{xue2015does} explore the theoretical explanations for re-balancing
imbalanced data. Hsu et al.~\cite{ming2015unsupervised} propose a Closest
Common Space Learning (CCSL) algorithm by exploiting both label and structural
information for data within and across domains. This is achieved by learning
data correlations~\cite{ma2016decorrelation} and related latent source-target
domain pairs. RecSys-DAN is similar to CCSL, but it distinguishes itself by
integrating representation learning and adversarial learning in recommender
system domain. While the typical cross-domain recommendation is in line with data
imbalance problem, RecSys-DAN aims to transfer knowledge from a domain with
abundant data to a domain with scarce data instead of directly re-balancing
data.

\subsection{Generative Adversarial Network (GANs)}

Generative Adversarial Network (GANs)~\cite{goodfellow2014generative} is the
most successful method in adversarial learning. Recently, many GAN-based
extensions are proposed in different areas: image generation (e.g.,
DCGAN~\cite{radford2015unsupervised} and Wasserstein
GAN~\cite{arjovsky2017wasserstein}), NLP (e.g., SeqGAN~\cite{yu2017seqgan}
% and neural dialogue generation~\cite{li2017adversarial}
) and domain transfer
problem~\cite{tzeng2017adversarial}. In recommender systems
community, IRGAN \cite{wang2017irgan} is the first work to integrate GANs into
item-based recommendation. Differently, RecSys-DAN can be viewed as the first
work which explores the power of GAN in the context of cross-domain
recommender systems.

\subsection{Domain Adaptation}

Transfer learning~\cite{li2009transfer,pan2013transfer} has been
recently proposed to address the data sparsity problem in recommender
systems~\cite{pan2016survey,zhao2013active}. Domain adaptation, as a special
form of transfer learning, arises with the hypothesis that large amounts of
labeled data from a source domain are somehow similar to that in the unlabeled
target domain. It has been applied to learn domain transferable representation
in a variety of computer vision tasks~\cite{ganin2015unsupervised, sener2016learning, taigman2016unsupervised,
tzeng2017adversarial, xu2018cross}. Domain-Adversarial Neural
Network (DANN)~\cite{ganin2015unsupervised} learns domain-invariant features
with adversarial training. Domain Transfer Network
(DTN)~\cite{taigman2016unsupervised} translates images across domains. E.
Tzeng et al. propose a unified framework, Adversarial Discriminative Domain
Adaptation (ADDA)~\cite{tzeng2017adversarial}, for object classification task.
RecSys-DAN is partly inspired by ADDA, though there are many differences
between ADDA and RecSys-DAN. RecSys-DAN is different to existing adaptation
methods mainly in two aspects: RecSys-DAN adopts multi-level generators and
discriminator for user/item features and their interactions, and it can
captures features from multimodal data~\cite{WangMM}.
\newcommand*{\Scale}[2][4]{\scalebox{#1}{$#2$}}%
\section{Motivation and Problem Statement}
\label{sec:motivation}
\begin{figure*}
    \begin{subfigure}[t]{0.245\textwidth}
        \includegraphics[width=4.5cm, height=2.25cm]{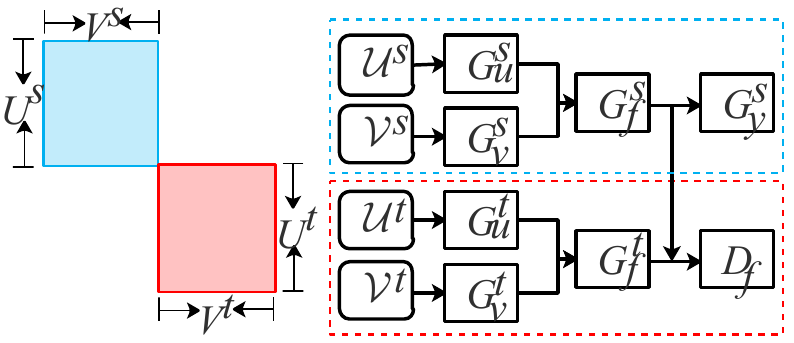}
        \caption{UI-DAN}
        \label{fig:framework:a}
    \end{subfigure}
    \begin{subfigure}[t]{0.24\textwidth}
        \includegraphics[width=4.3cm, height=2.25cm]{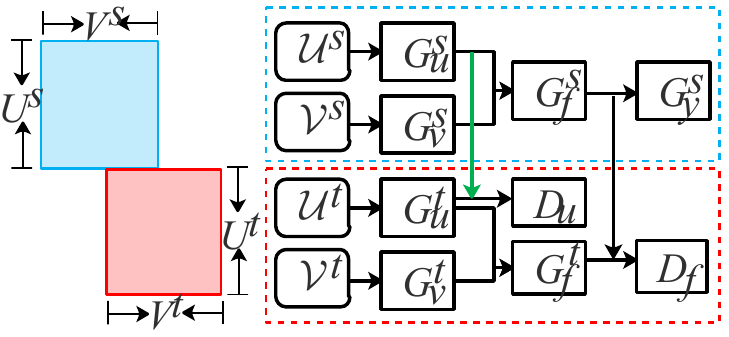}
        \caption{U-DAN}
        \label{fig:framework:b}
   \end{subfigure}
    \begin{subfigure}[t]{0.25\textwidth}
        \includegraphics[width=4.6cm, height=2.25cm]{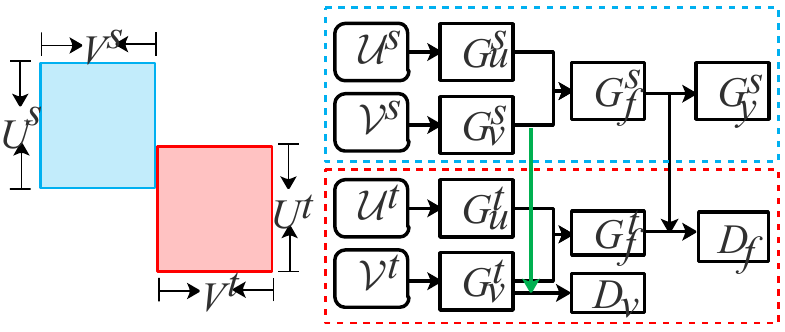}
        \caption{I-DAN}
        \label{fig:framework:c}
    \end{subfigure}
    \begin{subfigure}[t]{0.245\textwidth}
        \includegraphics[width=4.2cm, height=2.25cm]{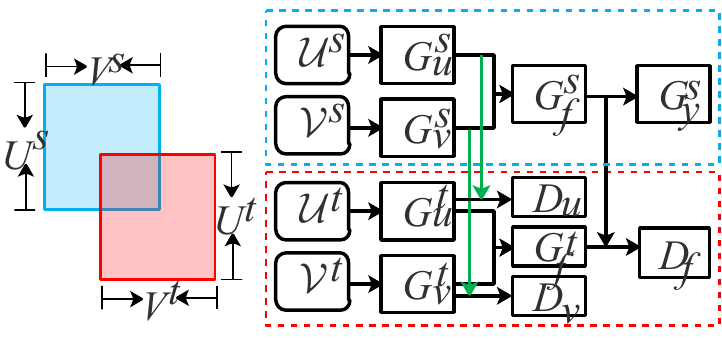}
        \caption{H-DAN}
        \label{fig:framework:d}
    \end{subfigure}
\caption{RecSys-DAN instantiations. $U^k$, $V^k$, $k\in\{s,t\}$ are user and item sets in domain $k$. The overlaps show that the shared user set of $U^s$ and $U^t$, or shared item set of $V^s$ and $V^t$ . $G_u$, $G_v$, $G_f$ ($D_u$, $D_v$, $D_f$) are corresponding to user, item and interaction feature generators (discriminators). The goal is: learning to align the latent representations between a source domain and a target domain that discriminators cannot distinguish. $G_y^s$ is the scoring function in the source domain.}
\label{fig:framework}
% \vspace{-5pt}
\end{figure*}

\vspace{5pt}
\subsubsection{Motivation} Motivated by the success of GANs and domain adaptation, RecSys-DAN aims to
address the data sparsity and data imbalance problem in a target domain by
adapting the object (user or item) and their interactions from a source
domain, i.e., learning to align user, item and user-item preference
representations across domains via discriminative adversarial domain
adaptation.

\vspace{5pt}
\subsubsection{General Problem} We first formalize the typical setting of a recommender system. Let $\mathcal{D}$ be a dataset consisting of $N$ users $U=\{\mathcal{U}_1,...,\mathcal{U}_N\}$ and $M$ items $V=\{\mathcal{V}_1,...,\mathcal{V}_M\}$. The user-item preferences can be represented as a rating matrix $\mathcal{Y}\in \mathbb{R}^{N\times M }$, where $\mathcal{Y}_{uv}$ is user $\mathcal{U}$'s preference rating on item $\mathcal{V}$. 
We denote by $\mathcal{U}=V(\mathcal{U})=\{\mathcal{V} \in V| \mathcal{Y}_{uv}\neq  0\}$, the set of items on which user $\mathcal{U}$ has non-zero preference values. Similarly, we use $\mathcal{V}=U(\mathcal{V})=\{\mathcal{U} \in U| \mathcal{Y}_{uv}\neq  0\}$ to indicate the set of users who have non-zero ratings on item $\mathcal{V}$. The task of recommender systems is to learn a function $h$ to predict the preference rating $\hat{\mathcal{Y}}_{uv}$ of user $\mathcal{U}$ for item $\mathcal{V}$ so that $\hat{\mathcal{Y}}_{uv}$ approximates ground-truth preference score $\mathcal{Y}_{uv}$. The function $h$ often has the following form:
\begin{equation}
\label{eq:scoring}
\hat{\mathcal{Y}}_{uv}=h(\mathcal{U},\mathcal{V}; \Theta_h),
\end{equation}
where $\Theta_h$ are the learnable parameters of $h$. The users and items are associated with existing features such as product metadata when available. The denser the user-item preference matrix $\mathbf{P}$ is, the less challenging the learning and prediction problems are. 
However, $\mathbf{P}$ can be very sparse in practice. 

\vspace{5pt}
\subsubsection{Adversarial Cross-Domain Alignment} To address this type of data
sparsity problem, we propose to perform domain adaptation going from a
\textit{source} domain with several user-item preference values to a
\textit{target} domain with \emph{no} user-item preferences.  Specifically,
the proposed approach learns a function $G$ that maps the following objects to latent vector representations:
the set of items that represented as $\mathcal{U}$;  the set of users that represented as $\mathcal{V}$; the set of user-item pairs ($\mathcal{U}, \mathcal{V}$). The $G$ is learned in a way that a discriminator $D$
cannot distinguish the latent representations generated for the target domain
from the latent representations generated for the source domain. We achieve
this by introducing an adversarial learning loss involving $G$ and $D$. For
the sake of readability, we refer to $G$ as a generator and write $G^{k}_{j}$
to denote different types of generators with $k \in \{s, t\}$ (source or
target) and $j \in \{u, v, f\}$ (user, item or item-user pairs).

Contrary to existing work, we formulate the adversarial loss for different
types of objects (users and items) and their interactions. The adversarial
loss, therefore, aligns distributions of latent items and user representations
\emph{as well as} their relationships given by the user-item preferences. The
latent representations computed by the generators, therefore, fall into three
categories: (1) user representations; (2) item representation;
and (3) interaction representations of user-item pairs.

\vspace{5pt}
\subsubsection{Shared Cross-Domain Objects}  Learning across domains
requires the existence of some relations in the participating domains.
Usually, this relation is formed when objects (users,  items) are found to be
common in both domains \cite{khan2017cross}. 
To cover the different scenarios, RecSys-DAN includes four different adversarial cross-
domain adaptation scenarios as below. They are classified according to whether a subset of user set
$U$ and item set $V$ exists in both source and target domains:

\begin{itemize}
\item Interaction adaptation: ~~~~~~$U^s \cap U^t=\varnothing$ and $V^s\cap V^t=\varnothing$.   
\item User adaptation: ~~~~~~~~~~~~~~~$U^s \cap U^t = \varnothing$ and $V^s\cap V^t\neq\varnothing$.
\item Item adaptation: ~~~~~~~~~~~~~~~$U^s \cap U^t \neq \varnothing$ and $V^s\cap V^t=\varnothing$.
\item Hybrid adaptation: ~~~~~~~~~~~$U^s \cap U^t \neq \varnothing$ and $V^s\cap V^t\neq\varnothing$.
\end{itemize}

Correspondingly, we proposed UI-DAN, U-DAN, I-DAN and H-DAN as shown in Fig.~\ref{fig:framework}. The additional discriminators (in green) are
introduced for shared objects. For instance, in the user adaptation scenario
(U-DAN) where the set of users in the source and target domain are disjoint,
we introduce a discriminator $D_u$ attempting to distinguish between latent
user representations from the source and target domain in order to align those
representations in latent space.  
\section{Discriminative Adversarial Networks for Cross-Domain Recommendation}
\label{sec:method}

We firstly describe the learning of representations of
objects (i.e., user and item) and their interactions. Then we elaborate
on the objectives of learning to align the representations across domains.
Finally, we introduce RecSys-DAN as a generalized adversarial adaptation
framework.

\subsection{Learning Domain Representations}

Given a set of users, items and ratings in the source domain, we can learn a
latent representation space $\mathcal{X}^s\in \mathbb{R}^d$ by computing a
supervised loss on the given input $X^s$ and ratings $Y^s$. Since
the target domain has no or only few ratings, we do not directly learn the
representations for the target domain. Instead, we learn mappings from the
source representation space $\mathcal{X}^s$ to the target representation
space $\mathcal{X}^t\in \mathbb{R}^d$ so as to minimize the distance between
them. This can be achieved by first parameterizing source and target mapping
functions, $M^s: X^s\rightarrow \mathcal{X}^s$ and $M^t: X^t\rightarrow
\mathcal{X}^t$, and then minimizing the distance between the empirical source
and target mapping distributions: $M^s(X^s)$ and
$M^t(X^t)$~\cite{tzeng2017adversarial}.  In this work, $M^k=\{G_u^k,
G_v^k,G_f^k\}, k \in {\{s, t\}}$ is a set consisting of user mapping function $G_u^k$ and item mapping function
$G_v^k$, and user-item pair mapping function $G_f^k$.

For learning textual representations, the $G_u^k$ is a recurrent neural network (RNN), specifically, RecSys-DAN adopts Long Short-Term Memory (LSTM)~\cite{HochreiterS97}:
\begin{align}
\begin{split}
\label{equ:input}
& i_t=\sigma (W_{xi}x_t+W_{hi}h_{t-1}+b_i)\\
& f_t=\sigma (W_{xf}x_t+W_{hf}h_{t-1}+b_f)\\
& o_t=\sigma (W_{xo}x_t+W_{ho}h_{t-1}+b_o)\\
& g_t=\tanh (W_{xc}x_t+W_{hc}h_{t-1}+b_c)\\
& c_t=f_t \odot c_{t-1}+i_t\odot g_t\\
& h_t=o_t\odot\tanh(c_t)
\end{split}
\end{align}
where $i_t$, $f_t$ and $o_t$ are input, forget and output gate respectively, $c_t$ is memory cell. 
$G_v^k$ can be either RNN-based (when review texts are used to represent an item) or convolutional neural network (CNN)-based for visual representations (when product image is used to represent an item).  As shown in Fig.~\ref{fig:framework}, the mapping function $G_u^k (\mathcal{U}^k; \Theta_u^k):
\mathcal{U}^k  \rightarrow \mathcal{X}^k_u\in\mathbb{R}^d$ that maps a user
sample to a $d$ dimensional vector  $\mathcal{X}^k_u$ and is parameterized by
$\Theta_u^k$. Similarly, we have the item mapping function $G_v^k
(\mathcal{V}^k; \Theta_v^k): \mathcal{V}^k  \rightarrow
\mathcal{X}^k_v\in\mathbb{R}^d$. Given user and item representations
($\mathcal{X}_u^k, \mathcal{X}_v^k$), mapping function
$G_f^k(\mathcal{X}^k_u, \mathcal{X}^k_v; \Theta_f^k):(\mathcal{X}^k_u,
\mathcal{X}^k_v) \rightarrow \mathcal{X}^k_f\in\mathbb{R}^d$ learns user-item
interaction representation $\mathcal{X}^k_f$. The prediction
$\hat{\mathcal{Y}}=h^k\Big(G_f^k(\mathcal{X}^k_u, \mathcal{X}^k_v;
\Theta_f^k);\Theta_h^k\Big)$, where $h^k$ is the scoring function in
Eq.~\ref{eq:scoring}. Since there in only one scoring function can be learned in supervised way, i.e. $G_y^s$,  and $h^s=h^t=G_y^s$, we use $G_y^s$ to represent the scoring function. 
In a source domain,  the parameters $\Theta^s=\{\Theta_u^s, \Theta_v^s,
\Theta_f^s, \Theta_h^s\}$ are learned by optimizing the objective:
\begin{equation}
\min_\mathbf{\Theta^s}\left [\frac{1}{\left |\mathcal{D}  \right |}\sum_{i=1}^{|\mathcal{D}|}\mathcal{L}^s(\mathcal{U}_i^s, \mathcal{V}_i^s, \mathcal{Y}_i^s )+\lambda \left \| \Theta^s  \right \|\right ]
\label{equ:s_loss}
\end{equation}
where $\langle$$\mathcal{U}_i^s$, $\mathcal{V}_i^s$,
$\mathcal{Y}_i^s$$\rangle$ presents raw $\langle$user, item, truth
score$\rangle$ triple, and $\mathcal{L}^s(\mathcal{U}_i^s,
\mathcal{V}_i^s,\mathcal{Y}_i^s)=\parallel  \hat{\mathcal{Y}}^s_i-\mathcal{Y}_i^s\parallel ^2$.
$\left |\mathcal{D}  \right |$ is the size of training set. $\lambda$ is the
regularization parameter. By minimizing the objective function
(\ref{equ:s_loss}), the mapping functions $G_u^s$, $G_v^s$ and $G_f^s$ can be
learned and used for extracting user, item and user-item features
respectively in source domain by fixing corresponding parameter. For the
unlabeled target domain, the corresponding target mapping functions $G_u^t$,
$G_v^t$ and $G_f^t$ can be learned adversarially as we will explain in the
next section.

\subsection{Adversarial Representation Adaptation}

One of the algorithmic principles of domain adaptation is to learn a space in
which source and target domains are close to each other while keeping good
performances on the source domain task~\cite{pan2010survey}. Following the
settings of standard GAN~\cite{goodfellow2014generative}, domain
discriminators $D_u$, $D_v$ and $D_f$ in RecSys-DAN are designed to perform
min-max games and adversarially learn target generators (i.e., mapping
functions) $G_u^t (\mathcal{U}^t; \Theta_u^t)$, $G_v^t (\mathcal{V}^t;
\Theta_v^t)$ and $G_f^t(\mathcal{X}_u^t, \mathcal{X}_v^t;
\Theta_f^t)$ with unlabeled samples. The loss functions of each
instantiation of RecSys-DAN are as follows:
\begin{itemize}
\setlength\itemsep{3pt}
\item  UI-DAN: $\Scale[1]{ \min\limits_{G_f^t}\max\limits_{D_f}\mathcal{L}(D_f, G_f^t)}$\\  $\Scale[1]{\ \ \ \ \ \ \ \ \ \ \ \ s.t. ~U^s\cap U^t=\varnothing$ and $V^s\cap V^t=\varnothing}$.   
\item  U-DAN: $\Scale[1]{ \min\limits_{G_f^t, G_u^t }\max\limits_{D_f, D_u}\mathcal{L}(D_f, D_u, G_f^t, G_u^t)}$\\ $\Scale[1]{\ \ \ \ \ \ \ \ \ \ \ \ s.t. ~ U^s\cap U^t = \varnothing$ and $V^s\cap V^t\neq\varnothing}$
\item  I-DAN: $\Scale[1]{\min\limits_{G_f^t, G_v^t }\max\limits_{D_f, D_v}\mathcal{L}(D_f, D_v, G_f^t, G_v^t,)}$\\ $\Scale[1]{\ \ \ \ \ \ \ \ \ \ \ \ s.t. ~ U^s\cap U^t \neq \varnothing$ and $V^s\cap V^t=\varnothing}$
\item  H-DAN: $\Scale[1]{\min\limits_{G_f^t, G_u^t , G_v^t }\max\limits_{D_f, D_u, D_v}\mathcal{L}(D_f, D_u, D_v, G_f^t, G_u^t , G_v^t)}$ \\ $\Scale[1]{\ \ \ \ \ \ \ \ \ \ \ \ s.t. ~ U^s\cap U^t \neq \varnothing$ and $V^s\cap V^t\neq\varnothing}$
\end{itemize}

The objectives are learning generators in the target domain to generate features
$\mathcal{X}^t\in\{\mathcal{X}_u^t, \mathcal{X}_v^t, \mathcal{X}_f^t\}$ which
are intended to be close to the source latent representations
$\mathcal{X}^s\in \{ \mathcal{X}_u^s, \mathcal{X}_v^s, \mathcal{X}_f^s\}$.
More specifically, $G_f$ generates interaction-level domain indistinguishable
features, while $G_u$/$G_v$ generates indistinguishable user/item features for overlapping users/items.
Formally, the source generators  $M^s=\{G_f^s, G_u^s, G_v^s\}$ and predictor
$G_y^s$ is learned in a supervised way:
\begin{align}
\begin{split}
\min_{G_y^s, M^s}\mathcal{L}_s(U^s, V^s,Y^s) \\
 &\hspace{-25mm}= \mathbb{E}_{{(\mathcal{U}^s, \mathcal{V}^s,\mathcal{Y}^s)} \sim (U^s, V^s, Y^s)}[(G_y^s(M^s, \mathcal{U}^s, \mathcal{V}^s,\mathcal{Y}^s)] \\
&\hspace{-25mm}=\frac{1}{\left |\mathcal{D}^s  \right |}\sum_{i=1}^{|\mathcal{D}^s|}\mathcal{L}^s(\mathcal{U}_i^s, \mathcal{V}_i^s,\mathcal{Y}_i^s)+\lambda \left \| \Theta^s  \right \| \\
&\hspace{-25mm}=\frac{1}{\left |\mathcal{D}^s  \right |}\sum_{i=1}^{|\mathcal{D}^s|}(\hat{\mathcal{Y}}^s_i-\mathcal{Y}_i^s )^2+\lambda \left \| \Theta^s  \right \|
\end{split}
\end{align}

The optimization of source weights $\Theta^s$ is formulated as a regression task which
minimizes the mean squared error (MSE) over samples. In learning target
generators $M^t=\{G_f^t, G_u^t, G_v^t\}$, $M^s$ is used as a domain
regularizer with fixed parameters. This is similar to the original
GAN~\cite{goodfellow2014generative} where a generated space is updated with a
fixed real space. To simplify, we take UI-DAN as an exemplary illustration,
the learning objective is:
\begin{align}
\begin{split}
\max_{D_f}\mathcal{L}_f(U^s, V^s, U^t, V^t, M^s, M^t)\\
&\hspace{-40mm}=\mathbb{E}_{{(\mathcal{U}^s, \mathcal{V}^s)} \sim (U^s_u, V^s_v)}[\log D_f(M^s(\mathcal{U}^s, \mathcal{V}^s))]\\
&\hspace{-40mm} + \mathbb{E}_{{(\mathcal{U}^t, \mathcal{V}^t)} \sim (U^t_u, V^t_v)}[\log(1-D_f(M^t(\mathcal{U}^t, \mathcal{V}^t)))]
\end{split}
\end{align}
\begin{align}
\begin{split}
\min_{M^t}\mathcal{L}_{m}(U^t, V^t, D_f)\\
&\hspace{-20mm}=\mathbb{E}_{{(\mathcal{U}^t, \mathcal{V}^t)} \sim (U^t, V^t)}[\log(1-D_f(M^t(\mathcal{U}^t, \mathcal{V}^t)))]
\end{split}
\end{align}
where $M^t$ is initialized with $M^s$.

With learned $M^t$, user, item, interaction representations $\mathcal{X}_u^t$, $\mathcal{X}_v^t$, $\mathcal{X}_f^t$ can be extracted as inputs for scoring function $G_y^s$, which makes
preference predictions. Note that one of the essential differences between RecSys-DAN
and prior recommendation methods is that ResSys-DAN takes the cross-domain
overlap users (items) into account to learn indistinguishable user (item)
representation as shown in Fig.~\ref{fig:framework:b}, Fig.~\ref{fig:framework:c}, and Fig.~\ref{fig:framework:d}. With shared users and
items across domain, additionally, $D_u$ and $D_v$ are designed and lead to
scenarios that have interaction-level $D_f$, $G_f$ and feature-level $D_u$,
$D_v$, $G_u$, $G_v$:
\begin{align}
\begin{split}
\max_{D_z}\mathcal{L}_z(U^s, V^s, U^t, V^t, G_u^s, G_v^s, G_u^t, G_v^t) \\
\min_{G^t_z}\mathcal{L}_{m}(U^t, V^t, D_z)\\ 
s.t. ~~D_z=D_u~~\text{if}~ U^s \cap U^t = \varnothing~\text{and} ~V^s\cap V^t\neq \varnothing \\
~~ D_z=D_v~~\text{if}~U^s \cap U^t \neq \varnothing~\text{and} ~V^s\cap V^t = \varnothing  \\
~~D_z=D_u,D_v~~\text{if}~ U^s \cap U^t \neq \varnothing~\text{and} ~V^s\cap V^t\neq \varnothing
\end{split}
\end{align}
The optimization of the additional discriminators and generators is achieved by fine-tuning $G_u^t$ ($G_v^t$) on cross-domain shared user/item subset.

\begin{algorithm}
 \small
  \caption{Learning algorithm for UI-DAN}
  \label{alg}
    \KwIn{source set $\mathcal{D}^s=\{X_u^s,X_v^s, Y^s\}$, 
           target set $\mathcal{D}^t=\{X_u^t, X_v^t\}$, dummy domain label $Y^d\in\{0,1\}$, batch size $\mathcal{B}$.}
  	\textbf{Initilize: $M^s, M^t, G_y^s, D_f$} \\
    $\mathcal{N}^s=|\mathcal{D}^s|$,  $\mathcal{N}^t=|\mathcal{D}^t|$ \\
 	 \textit{pre-train on source domain:} \\
 	\Repeat{\text{stopping criterion is met}}{
\For {$b\leq \frac{\mathcal{N}^s}{\mathcal{B}}$}
  { mini batch $(\mathcal{U}_b^s, \mathcal{V}_b^s, \mathcal{Y}_b^s) \in (X_u^s, X^s_v, Y^s)$ \\
  
 $M^s, G_y^s \Leftarrow  \min \mathcal{L}_s(\mathcal{U}_b^s, \mathcal{V}_b^s,\mathcal{Y}_b^s)$ 
}
}
  \textit{train generators on target domain}: \\
set $M^t\Leftarrow M^s$, and fix $M^s$  \\
 \Repeat{\text{stopping criterion is met}}{ 
 \For {$b\leq \frac{\mathcal{N}^s}{\mathcal{B}}$}
  {
    mini batch $(\mathcal{U}_b^s, \mathcal{V}_b^s) \in (X_u^s, X^s_v)$ \\
    \For {$k\leq \frac{\mathcal{N}^t}{\mathcal{B}}$}
    {
    mini batch $(\mathcal{U}_k^t, \mathcal{V}_k^t) \in (X_u^t, X_v^t)$ \\
	$D_f\Leftarrow  \max \mathcal{L}_f(\mathcal{U}_b^s, \mathcal{V}_b^s,\mathcal{U}_k^t, \mathcal{V}_k^t,\mathcal{Y}^d)$ \\   
	$M_t\Leftarrow  \min \mathcal{L}_m(\mathcal{U}_k^t, \mathcal{V}_k^t)$ \\   
}
 }
 }
 
\KwOut{$M^t$}    
 \textit{inference on target domain}: \\
$\hat{y}_t \Leftarrow G_y^s(M^t(x_u^t, x_v^t))$
\end{algorithm} 
\vspace{-2mm}

\subsection{Generalized Framework}
RecSys-DAN is a generalized framework. The choice of RecSys-DAN
instantiations is based on considering the following questions: (1) Which
type of modalities (e.g. numerical rating, review or image) are used to represent
$\mathcal{U}$ and $\mathcal{V}$? (2) Are there shared users and/or items
across domains? (3) Which adversarial objective is used?

The training procedure of each instantiation is different to each other, but
they also share some similarities. Algorithm 1 summaries the learning
procedure of UI-DAN in which two training stages are involved.  First, the pre-training in
the source domain for obtaining source generators $M^s$ and scoring
function $G_y^s$. The update of parameters $\Theta_u^s, \Theta_v^s, \Theta_f^s$ are achieved by: 
\begin{align}
\small
\begin{split}
\Theta_j^s:=\Theta_j^s -\eta \nabla_{\Theta_j^s}\frac{1}{\mathcal{B}}\sum_{i=1}^\mathcal{B}\mathcal{L}_s(\mathcal{U}_i^s, \mathcal{V}_i^s,\mathcal{Y}_i^s),~~j\in\{u,v,f\} 
\end{split}
\end{align}
where $\mathcal{B}$ is a min-batch of training samples, $\eta$ is learning rate.
Similarly, the optimal weights for scoring function $G_y^s(G_f^s;
\Theta_y^s)$ can be learned.  Second, cross-domain adversarial learning, the goal is to learn the
target generators $M^t$ in an adversarial way. By using dummy domain labels,
$y^d=1$ presents the data from source domain and $y^d=0$ for target domain.
The domain discriminator $D_f(G_f^s, G_f^t; \Theta_d)$ is obtained by
ascending stochastic gradients~\cite{goodfellow2014generative} at each batch
using the following update rule:
\begin{align}
\Theta_d:= \Theta_d+\eta \nabla_{\Theta_d}\frac{1}{\mathcal{B}}\sum_{i=1}^\mathcal{B}\mathcal{L}_f(\mathcal{U}_i^s, \mathcal{V}_i^s,\mathcal{U}_i^t, \mathcal{V}_i^t,\mathcal{Y}_i^d)
\end{align}

Note that target generators $M^t$ is initialized with and updated in similar
way as $M^s$. By doing this, $M^t$ tries to push the user-item interaction
representations in the target domain as close as possible to the source
domain. Additionally, the ratings (i.e., labels) in the target domain are
never accessed in learning procedures of RecSys-DAN. As a comparison,
existing recommendation methods fail to handle this scenario. With learned
$M^t$, the rating regression can be performed with source score function
$G_y^s$ for a given user-item pair in the target domain:
\begin{align}
\hat{\mathcal{Y}}_{uv} \Leftarrow G_y^s\big(M^t(\mathcal{U}^t, \mathcal{V}^t)\big).
\end{align}
The learning procedures of U-DAN, I-DAN and
H-DAN have additional fine-tuning stage with training samples of shared
users/items. 
\vspace{-2mm}
\begin{algorithm}
 \small
  \caption{Learning for U-DAN and I-DAN}
  \label{alg1}
    \KwIn{$\mathcal{D}^s=\{X_u^s,X_v^s, Y^s\}$,  $\mathcal{D}^t=\{X_u^t, X_v^t\}$, \\
           shared item set $\mathcal{D}^o_u=\{X_v^o, X_u^s,X_u^t\}$,\\
           shared user set $\mathcal{D}^o_v=\{X_u^o, X_v^s,X_v^t\}$, $Y^d\in\{0,1\}$.}
  	\textbf{Initialize: $M^s, M^t, G_y^s, D_u,D_v,D_f$} \\
 	 \textit{call Algorithm \ref{alg} to obtain $M^t$, learning rate $\eta \times 0.001$} \\
  \textit{learning U-DAN}: \\
 \Repeat{\text{stopping criterion is met}}{ 
 \For {each batch $b$, $(\mathcal{V}_b^o, \mathcal{U}_b^s,\mathcal{U}_b^t) \in (X_v^o,X_u^s, X^t_u)$}
  {
	$D_u^t\Leftarrow  \max \mathcal{L}_f(\mathcal{V}_b^o, \mathcal{U}_b^s,\mathcal{U}_b^t,\mathcal{Y}^d_b)$ \\   
	$G_u^t\Leftarrow  \min \mathcal{L}_m(\mathcal{V}_b^o,\mathcal{U}_b^t)$ \\  
 }
  }
   \textit{learning I-DAN}: \\
 \Repeat{\text{stopping criterion is met}}{ 
 \For {each batch $b$, $(\mathcal{U}_b^o, \mathcal{V}_b^s,\mathcal{V}_b^t) \in (X_u^o, X_v^s, X^t_v)$}
  {
	$D_v^t\Leftarrow  \max \mathcal{L}_f(\mathcal{U}_b^o, \mathcal{V}_b^s,\mathcal{V}_b^t,\mathcal{Y}^d_b)$ \\   
	$G_v^t\Leftarrow  \min \mathcal{L}_m(\mathcal{U}_b^o,\mathcal{V}_b^t)$ \\  
 }
 }
\KwOut{$G_u^t, G_v^t$}    

\end{algorithm} 
\vspace{-2mm}
Algorithm \ref{alg1} presents the learning for  U-DAN and I-DAN while H-DAN is a combination of them. 
\section{Experiments}
\label{sec:exp}
This section evaluates the performance of RecSys-DAN on both unimodal and multimodal scenarios.

\begin{table}[!t]
\footnotesize
\centering
\caption{Overview of the datasets ($\dagger$ presents training samples for shared users and items respectively)}
\label{tab:dataset}
\begin{tabular}{|l|p{0.75cm}lp{1.8cm}|p{0.85cm}|}
 \hline
$D^s \rightarrow D^t$ & User  & Item & Sample     & $\mathcal{|VOC|}$                                      \\ \hline\hline
DM                    & 5540  & 3558 & 64544      & \multicolumn{1}{l|}{\multirow{3}{*}{4696}} \\ 
MI                    & 1429  & 891  & 10156      & \multicolumn{1}{l|}{}                      \\ 
DM $\cap$ MI               & 23    & 0    & 23         & \multicolumn{1}{l|}{}                      \\ \hline
HK                    & 14285 & 3227 & 41810      & \multirow{3}{*}{3651}                      \\ 
OP                    & 4773  & 1312 & 28044      &                                            \\ 
HK $\cap$ OP              & 1709  & 0    & 1709       &                                            \\ \hline 
CDs                   & 41437 & 9650 & 84432      & \multirow{3}{*}{10355}                     \\ 
DM                    & 5540  & 3558 & 6615       &                                            \\ 
CDs $\cap$ DM               & 4394  & 829  & 19529/6216$^\dagger$ &                                            \\ \hline
\end{tabular}
 \end{table}

\begin{table*}[!t]
\footnotesize
\caption{\small \label{tab:comparison_S_H} The results for UI-DAN and I-DAN in the unimodal and multimodal settings (s: source-only, a: adaptation, u: unimodal, m: multimodal). The best (supervised) baselines are in {\color{blue!95}\textbf{blue}}, and the best unimodal (multimodal) results of RecSys-DAN are in {\color{OliveGreen!95}\textbf{green}} ({\color{red!95}\textbf{red}}) . $\Delta =  (2\left | S_{-}^{*}-S_{+}^{*} \right |)/(S_{-}^{*}+S_{+}^{*})$ presents the percentage differences between the best result of ours ($S_{-}^*$, in green) and that of baselines ($S_{+}^*$, in blue). It demonstrates how close the performance of (unsupervised) RecSys-DAN to the performance of (supervised) baselines.}
\label{tab:modality}
\centering
\begin{center}

\begin{tabular}{|l|cccc|ccc|}
\hline
 $\mathcal{D}^s\rightarrow\mathcal{D}^t$   & \multicolumn{2}{c}{DM$\rightarrow$MI } & \multicolumn{2}{c|}{ HK $\rightarrow$ OP}& \multicolumn{3}{c|}{ Target Domain Training Data} \\ 

Models     &   RMSE        &  MAE         &  RMSE        &  MAE   & Rating& Review& Image      \\ \hline \hline
Normal  & 1.165 $\pm$ 0.022      & 0.843 $\pm$ 0.025     & 1.194 $\pm$ 0.024      & 0.894 $\pm$ 0.023 & Yes  & No& No  \\ 
KNN  & 1.040 $\pm$ 0.000      & 0.709 $\pm$ 0.000     &   0.957 $\pm$ 0.000    &  0.710 $\pm$ 0.000 & Yes  & No& No   \\ 
NMF   & 0.922 $\pm$ 0.009      & 0.644 $\pm$ 0.007      &    0.866 $\pm$ 0.003   &  0.637 $\pm$ 0.005 & Yes  & No& No   \\ 
SVD++  &  0.891 $\pm$ 0.008    & 0.648 $\pm$ 0.006     &  \color{blue!95}\textbf{0.844 $\pm$ 0.002}    &  0.642 $\pm$ 0.002  & Yes  & No& No    \\ 
HFT   &  0.914 $\pm$ 0.000   &  0.704 $\pm$ 0.000     &  0.917 $\pm$ 0.000   &  0.735 $\pm$ 0.000  & Yes  & Yes& No  \\ 
DeepCoNN  & \color{blue!95}\textbf{0.868 $\pm$ 0.002}   & \color{blue!95}\textbf{0.599 $\pm$ 0.003}     &  0.875 $\pm$ 0.001  &  \color{blue!95} \textbf{0.634 $\pm$ 0.001} & Yes  & Yes& No  \\
\hline  \hline 
UI-DAN (s, u)  &  1.087$\pm$ 0.180      &  0.918$\pm$ 0.002      &  0.959$\pm$ 0.028      &   0.684$\pm$ 0.003  & No & Yes& No  \\ 
I-DAN (s, u)   &    1.052$\pm$ 0.220    &    0.884$\pm$ 0.264    &     0.957$\pm$ 0.033   &   0.684$\pm$ 0.002 &No & Yes& No    \\ 
UI-DAN (s, m)  &  1.043$\pm$ 0.056      &  0.879$\pm$ 0.089      &  1.037$\pm$ 0.008      &   0.875$\pm$ 0.011  & No & Yes& Yes  \\ 
I-DAN (s, m)  &   1.450$\pm$ 0.291    &   1.296$\pm$ 0.308    &     1.953$\pm$ 0.290   &  1.759$\pm$ 0.286 & No & Yes& Yes    \\ \hline  \hline 
UI-DAN (a, u)   & 0.920$\pm$ 0.223      & \color{OliveGreen}\textbf{0.674$\pm$ 0.021}      &  0.917$\pm$ 0.005     &    0.674$\pm$ 0.002& No & Yes& No    \\ % \hline  \hline
I-DAN (a, u)   &    \color{OliveGreen} \textbf{0.914$\pm$ 0.002}  &   0.675$\pm$ 0.021    &   \color{OliveGreen}\textbf{0.911$\pm$ 0.002}    &  \color{OliveGreen}\textbf{0.670$\pm$ 0.002} & No & Yes& No   \\  

UI-DAN (a, m)   &  \color{red!95}\textbf{0.991$\pm$ 0.077}     &  \color{red!95}\textbf{0.765$\pm$ 0.143}      &   \color{red!95}\textbf{0.934$\pm$ 0.004}     &     \color{red!95}\textbf{0.745$\pm$ 0.006} & No & Yes& Yes   \\  
I-DAN (a, m)   &    1.078$\pm$ 0.033  &   0.795$\pm$ 0.027    &   1.144$\pm$ 0.078   &   0.868$\pm$ 0.039  & No & Yes& Yes   \\ \hline 
$\Delta$   &   5.16\%$\pm$ 0.22\% &   11.78\%$\pm$ 1.88\%   &   7.64\%$\pm$0.23\%    &   5.52\% $\pm$0.23\%    & - & -& -\\ \hline
\end{tabular}
\end{center}
\end{table*}

\begin{figure*}[!]
\vspace{-5mm}
\centering
 \includegraphics[width=0.8\linewidth]{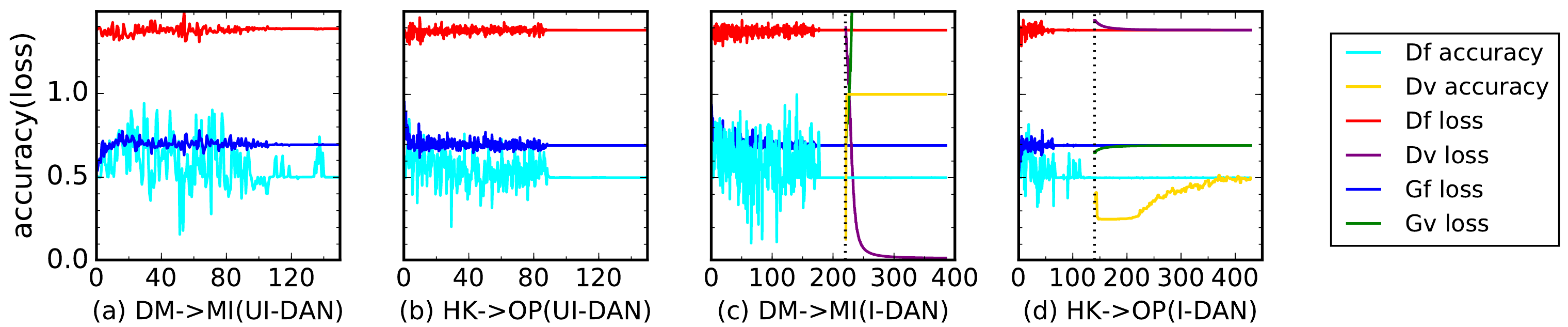}
\caption{\small Learning unimodal UI-DAN and I-DAN. It plots the changes of loss and accuracy of interaction-level (a-b) discriminator $D_f$/generator $G_f$ and item-level discriminator $D_v$/generator $G_v$ (c-d) on two dataset pairs against training epochs. The dash vertical lines in (c-d) denote the starting point for fine-tuning I-DAN. The X-axis presents the number of training epochs.}
\label{fig:unimodal_training}
\end{figure*}
\subsection{Dataset and Evaluation Metric}
\label{sec:data}

We evaluated RecSys-DAN on multiple sets on the Amazon
dataset~\cite{mcauley2015image}\footnote{\url{http://jmcauley.ucsd.edu/data/amazon/}},
which is widely used for evaluating recommender
systems~\cite{mcauley2013hidden,zheng2017joint}. It contains
different item and user modalities such as review text, product images and
ratings. We selected 5 categories to form three (source
$\rightarrow$ target) domain pairs: Digital Music$\rightarrow$Music
Instruments (DM$\rightarrow$MI), Home \& Kitchen$\rightarrow$Office Products
(HK$\rightarrow$OP) and CDs \& Vinyl$\rightarrow$Digital Music
(CDs$\rightarrow$DM). Some statistics of the datasets are listed in
Tab.~\ref{tab:dataset}.  $\mathcal{|VOC|}$  is the size of the vocabulary of
words used in reviews in the source and target training sets. Words which
occurred less than 5 times were removed. We randomly split each dataset into
80\%/10\%/10\% for training/validation/test. The training reviews
associated with a user/item were concatenated to present the user/item
following previous work~\cite{zheng2017joint}. We aligned users (items) that
occured in both the source and target domains to ensure an equal number of
training reviews for both domains.
We evaluated all the models on the rating prediction task using both the root
mean squared error (RMSE)  
and the mean average error (MAE):

\begin{equation}
\scriptsize
\text{RMSE}=\sqrt{\frac{1}{\mathcal{|D|}}\sum_{(\mathcal{U}, \mathcal{V})\in \mathcal{D}} ({\hat{\mathcal{Y}}_{uv}}-\mathcal{Y}_{uv})^2},\,\,\,\text{MAE}=\frac{1}{\mathcal{|D|}}\sum_{(\mathcal{U}, \mathcal{V})\in \mathcal{D}} |\hat{\mathcal{Y}}_{uv}-\mathcal{Y}_{uv}|
\end{equation}
% \begin{equation}
% \small 
% \text{MAE}=\frac{1}{\mathcal{|D|}}\sum_{(\mathcal{U}, \mathcal{V})\in \mathcal{D}} |\hat{\mathcal{Y}}_{uv}-\mathcal{Y}_{uv}|
% \end{equation}
where $\hat{\mathcal{Y}}_{uv}$ and $\mathcal{Y}_{uv}$ are predicted and truth rating, respectively.

\subsection{Baseline Methods}
We compare RecSys-DAN against a variety of methods.
Naive: \textbf{Normal} is a random rating predictor which gives predictions based on the (norm) distribution of the training set.
Matrix factorization: \textbf{NMF}~\cite{lee2001algorithms}, Non-negative Matrix Factorization that only uses ratings. And \textbf{SVD++}~\cite{koren2008factorization}, extended SVD for latent factor modeling.
Nearest neighbors: \textbf{KNN}~\cite{koren2010factor}.
Topic modeling: \textbf{HFT}~\cite{mcauley2013hidden}.
Deep learning methods: \textbf{DeepCoNN}~\cite{zheng2017joint}, which is the current state-of-the-art approach. Additionally, we compared RecSys-DAN with typical cross-domain recommendation methods.

Following previous work~\cite{sener2016learning,tzeng2017adversarial},
\textbf{source-only} results for applying a source domain models to the
target domain are also reported. Note that rating information in the target
domain is accessible to the baseline methods (except source-only), while
RecSys-DAN has no access to ratings in the target domain.

\subsection{Implementations}
We implemented RecSys-DAN with Theano\footnote{\url{http://www.deeplearning.net/software/theano/}}. The discriminators $D_f$, $D_u$, $D_v$ are formed
with following layers: Dense(512)$\rightarrow$Relu($\cdot$)$\rightarrow$Dense(2)$\rightarrow$Softmax($\cdot$). The architecture of generators varies
according to different scenarios. For unimodal scenario (textual user and
item representations), $G_u^s$, $G_v^s$, $G_u^t$, $G_v^t$ are formed by:
Embedding($\mathcal{|VOC|}$)$\rightarrow$LSTM (256)$\rightarrow$Average
Pooling, and $G_f^s$, $G_f^t$ are constructed using:
Dense(512)$\rightarrow$Dropout (0.5). For multimodal scenario
(textual user representation and visual item representation),
the main architecture of $G_v^s$, $G_v^t$ is: CNN$\rightarrow$Dense
(4096)$\rightarrow$ Dense(256), and other configurations remain unchanged as
in unimodal scenario. The weights of LSTM are orthogonally
initialized~\cite{saxe2013exact}. We used a batch size of 512. The models
were optimized with ADADELTA \cite{zeiler2012adadelta} and the initial
learning rate $\eta$ is 0.0001 (decreased by $\times$0.001 for
U-DAN, I-DAN and H-DAN). We implemented KNN, NMF and SVD++ using
SurPrise package\footnote{\url{http://surpriselib.com/}} and used authors'
implementations for
HFT\footnote{\url{http://cseweb.ucsd.edu/~jmcauley/code/code_RecSys13.tar.gz}}
and DeepCoNN\footnote{\url{https://github.com/chenchongthu/DeepCoNN}}. To make a fair comparison, implemented baselines are trained with grid search (for NMF and SVD++, regularization [0.0001, 0.0005, 0.001], learning rate [0.0005, 0.001, 0.005, 0.01]. For HFT, regularization [0.0001, 0.001, 0.01, 0.1, 1], lambda [0.1, 0.25, 0.5, 1]). For DeepCoNN, we use the suggested default parameters. The best scores are reported.

\subsection{Results and Discussions}
\label{sec:uni_modal}
We first evaluated two RecSys-DAN instances: UI-DAN (applied to the scenario
where source and target domains have neither overlapping users nor items) and
I-DAN (applied to the scenario where the source and target domains only shared some users) in the unimodal and multimodal scenarios. The results are
summarized in Tab.~\ref{tab:modality}.

\subsubsection{Unimodal RecSys-DAN} 
The results listed in Tab.~\ref{tab:modality} show that both UI-DAN and I-DAN improve the source-only
baselines. For instance, UI-DAN reduces the source-only error by $\sim$15\%
(RMSE) and $\sim$27\% (MAE) on DM$\rightarrow$MI. On HK$\rightarrow$OP, it
improves the source-only baselines by $\sim$4\% (RMSE) and $\sim$1.5\% (MAE),
respectively.  In the scenario where source and target domains share users,
I-DAN can improve UI-DAN on both dataset pairs ($\sim$0.4\% on average across
metrics). Compared to its source-only baselines, I-DAN achieves improvements
similar to those of UI-DAN.

Fig.~\ref{fig:unimodal_training}a and Fig.~\ref{fig:unimodal_training}b show
the changes of the loss/accuracy of the interaction discriminator $D_f$ and
the loss of $G_f$ against the number of epochs with the UI-DAN. On both
dataset pairs, the equilibrium points are reached at $\sim$100 epochs where
binary classification accuracy of discriminator is 50\%. It suggests that the
user-item interaction representation from generator is indistinguishable to
discriminator. When training I-DAN with shared user samples, we first trained
interaction-level $D_f$ and $G_f$ and then fine-tuned item-level $D_v$ and
$G_v$ by decreasing learning rate to 0.001$\times \eta$. 
We adopted small learning rate $\eta$ to ensure that $G_v$ could generate
indistinguishable item representation for shared users while maintaining
interaction-level representations. Figures~\ref{fig:unimodal_training}c
and~\ref{fig:unimodal_training}d present the training procedure of I-DAN. On
DM$\rightarrow$MI, $D_v$ and $G_v$ had difficulty to converge due
to limited shared user samples. On the contrary, with more shared samples, I-DAN was able to converge
on both interaction-level and item-level on HK$\rightarrow$OP. From
experimental results, we can observe that item-level representations are not
as important as interaction-level representation on rating prediction task.
Similar findings are reported in Tab.~\ref{tab:CDs}.

\subsubsection{Multimodal RecSys-DAN} 
\begin{figure}[!]
\vspace{-4mm}
\center
  \begin{subfigure}{0.75\linewidth}
    \centering
    \includegraphics[width=\linewidth]{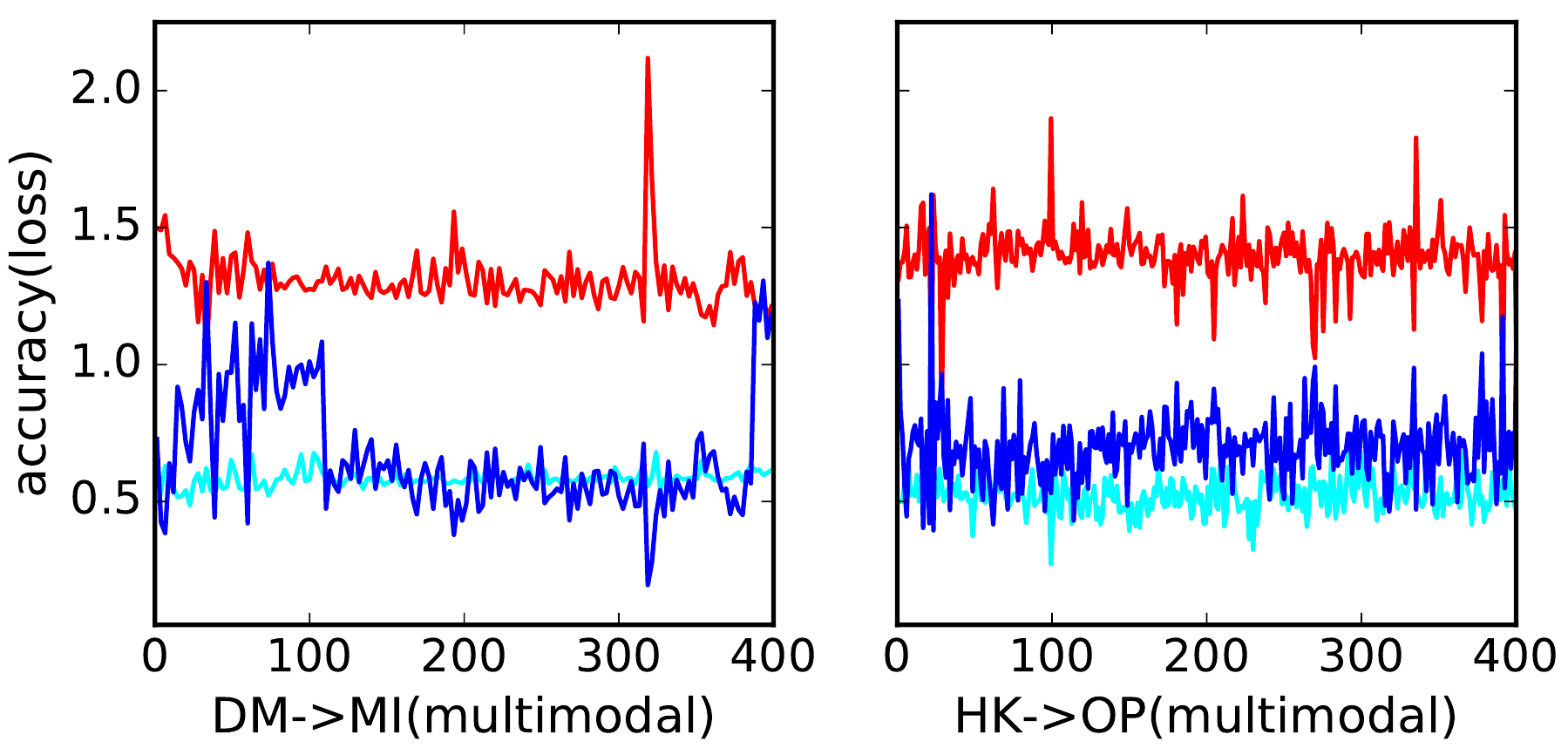}
  \end{subfigure}
\caption{Learning multimodal UI-DAN. The labels and legends are the same as Fig.~\ref{fig:unimodal_training}.}
\label{fig:multi_train}
\vspace{-10pt}
\end{figure}

The task becomes more challenging when both ratings and reviews are not
available. In this scenario, we replaced the review text of an item with its
image, if available, which leads to a multimodal unsupervised adaptation
problem. The correlations between textual user embeddings and visual item
embeddings need to be adapted across the given domains. The results of UI-DAN
(a, m) and I-DAN (a, m) in the multimodal settings can be found in
Tab.~\ref{tab:modality}. We find that it is more difficult to learn
user-item correlations across modalities, compared to the unimodal setting.
Fig.~\ref{fig:multi_train} presents the learning of multimodal adversarial
adaptation paradigm. Although the performance of multimodal UI-DAN and I-DAN
is not as good as the unimodal ones, it is still robust when addressing the
item-based cold-start recommendation problem. UI-DAN (a, m) and I-DAN (a, m),
however, significantly improve UI-DAN (s, m) and I-DAN (s, m). For instance,
I-DAN (a, m) outperforms I-DAN (s, m) by $\sim$26\% (RMSE)/$\sim$39\% (MAE)
for DM$\rightarrow$MI and $\sim$41\% (RMSE)/$\sim$51\% (MAE) for
HK$\rightarrow$OP, respectively.
\begin{table}
\footnotesize
\center
\caption{RecSys-DAN Results on CDs $\rightarrow$DM }
\label{tab:CDs}
\begin{tabular}{|l|cc|}
\hline
 $\mathcal{D}^s\rightarrow\mathcal{D}^t$   & \multicolumn{2}{c|}{CDs$\rightarrow$DM }  \\ 
Models     &   RMSE        &  MAE              \\ \hline \hline
Normal  & 1.452 $\pm$ 0.021     & 1.100 $\pm$ 0.022          \\ 
KNN  & 1.110 $\pm$ 0.000     & 0.870 $\pm$ 0.000         \\ 
NMF   & 1.062 $\pm$ 0.001      & 0.861 $\pm$ 0.001          \\ 
SVD++  &  1.061 $\pm$ 0.000    & 0.841 $\pm$ 0.001        \\ 
HFT   &  1.099 $\pm$ 0.000     &   0.869 $\pm$ 0.000      \\ 
DeepCoNN  & \color{blue!95}\textbf{1.038 $\pm$ 0.004}  &     \color{blue!95}\textbf{0.805 $\pm$ 0.003}       \\ 
\hline \hline
Source Only  &  1.131$\pm$ 0.028      & 0.857$\pm$ 0.080        \\ 
UI-DAN   &    1.076$\pm$ 0.002   &  0.791$\pm$ 0.019       \\ % \hline  \hline
U-DAN  &    1.071$\pm$ 0.005    &     0.784$\pm$ 0.002      \\ 
I-DAN  &  1.068$\pm$ 0.006    &   0.781$\pm$ 0.002     \\
H-DAN  &  \color{OliveGreen!95}\textbf{1.068$\pm$ 0.002}   &    \color{OliveGreen!95}\textbf{0.779$\pm$ 0.002}     \\ \hline  
$\Delta$   &    2.85\%$\pm$ 0.28\%   &  \textbf{3.28\%$\pm$ 0.32\%}  \\  \hline
\end{tabular}
\end{table}

\subsubsection{Compare Different Instances of RecSys-DAN}
An experiment was conducted on CDs$\rightarrow$DM (unimodal) where both
shared users and items existed to further explore the different instances of
RecSys-DAN. The results in Tab.~\ref{tab:CDs} illustrate that unsupervised
domain adaptation models improve source-only baseline by $\sim$4.8\% (RMSE)
and $\sim$7.7\% (MAE). We find that U-DAN, I-DAN and H-DAN did not bring
significant improvements over UI-DAN. This is similar to the results of I-DAN
and UI-DAN in Tab.~\ref{tab:modality}. We conjecture the main reason is that
the rating prediction task is primarily based on the user-item interactions
(e.g., users express preferences on items). The interaction representations
are therefore of crucial importance as compared to user-level and item-level
representations, though the shared users/items could be beneficial when connecting domains. % depends on domain shifts.

\subsubsection{Compare to Cross-domain Recommendation Models}
We now compare our proposed architectures with the state-of-the-art
supervised models. 
As the first attempt to utilize unsupervised adversarial domain adaptation
for the (cold-start) cross-domain recommendation, it is difficult to directly
compare RecSys-DAN with previous methods. Existing cross-domain (e.g.,
EMCDR~\cite{man2017cross}, CrossMF~\cite{iwata2015cross}, HST~\cite{liu2015non}) or hybrid collaborative filtering (e.g.,
DeepCoNN~\cite{zheng2017joint}, cmLDA~\cite{tan2014cross}) methods
are \textit{NOT} able to learn models in the scenarios where ratings and/or
review texts are completely not available for training. The Tab.~\ref{tab:dis} suggests previous methods' limitations, which are addressed by
our proposed adversarial domain adaptation method. Therefore, we compare
RecSys-DAN with supervised baselines indirectly. 
\subsubsection{Compare to Supervised Models}
We trained the baselines directly on the target domain with labeled samples (Normal, KNN, NMF and SVD++ were trained
with user-item ratings, while HFT and DeepCoNN were trained with both ratings
and reviews). The goal is to examine how close the performance of
unsupervised RecSys-DAN without labeled target data to those supervised
methods which can access labeled target data. The results are reported in
Tab.~\ref{tab:modality} and Tab.~\ref{tab:CDs}. By purely transferring the
representations learned in the source domain to the target domain, our
methods achieve competitive performance compared to strong baselines.
Specifically, RecSys-DAN is able to achieve similar performance as NMF and
SVD++ with unsupervised adversarial adaptation and it outperforms baselines
on MAE in Tab.~\ref{tab:CDs}. From the aforementioned analysis, we can conclude
that ResSys-DAN has much better generalization ability and it is more
suitable to address practical problems such as cold-start recommendation.

\begin{table}[!t]
\footnotesize
\center
\caption{The comparison with RecSys-DAN and existing cross-domain recommendation methods. Existing methods have difficulties in learning a recommendation model when ratings on the target domain are completely missing.}
\label{tab:dis}
\begin{tabular}{|l|lc|} \hline
Methods                 & Required Target Inputs &Target Learning \\ \hline \hline
EMCDR~\cite{man2017cross} & rating     & supervised                                                                 \\
DeepCoNN~\cite{zheng2017joint} & rating, review &supervised  \\
DLSCF~\cite{jiang2017deep} & rating, binary rating &supervised  \\
CrossMF~\cite{iwata2015cross} &    rating         &   supervised                                              \\   
%CF+FM   &   rating       &  supervised                                                     \\  
CTSIF\_SVMs~\cite{yu2018svms} &   rating       &  supervised                                                     \\  
HST~\cite{liu2015non} & ratings & supervised \\
cmLDA~\cite{tan2014cross}                       &  rating, review, description &    supervised           \\
RecSys-DAN        &     {review or image}     &        adversarial                                    \\       \hline
\end{tabular}
\vspace{-5mm}
\end{table}

\begin{table*}[!t]
\centering
\caption{\small Exemplary predictions of RecSys-DAN (UI-DAN) on the target test set of ``office product'' with HK$\rightarrow$OP cross-domain recommendation. The first two examples are unimodal and the last two examples are multimodal based prediction. The predictions are purely based on transferring the representations of user-item interaction in the source domain (``home \& kitchen'') via an unsupervised and adversarial way. ``$<$UNK$>$'' means the word is not included in built vocabulary dictionary $\mathcal{VOC}$. We removed punctuations in reviews.}
\label{tab:examplel} 
\scalebox{0.95}{
\begin{tabular}{|p{6.cm}|p{9cm}|p{1cm}p{1cm}|} 
\hlinewd{1pt}
Reviews written by user & Reviews and/or Images associated to item & Prediction& Truth  \\ \hline
 has four internal pockets which is a nice addition round rings but with the better $<$UNK$>$ closure handy but could use slight improvement (...)    &   just what we needed good item great organizer less useful than i thought although may be just right for some colorful organizing okay (...)   &    \textbf{4.58}       &   \textbf{5}   \\ \hline
worked well very cool great product great product works great awesome product well very easy to use  &   need a computer excellent for keeping organized in class durable easy to use super nice for presentations great quality and price great idea to (...)   &   \textbf{5.08}        &   \textbf{5}    \\  \hline
good tape $<$UNK$>$ not very good flow good boxes but they come $<$UNK$>$  & 
\begin{wrapfigure}{L}{0.07\textwidth} 
\vspace{-5mm}
\begin{minipage}{0.85\textwidth}
\includegraphics[width=0.1\linewidth, height=0.05\textheight]{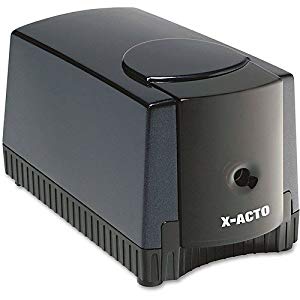}

\end{minipage} 
 \end{wrapfigure} 
 \vspace{-0.5mm}
efficient tool best value for price  while it lasted no frills sturdy sharpener  for frequent pencil $<$UNK$>$ sharp works as it should noisy but good excellent maybe not perfect for your use (...)

   &      \textbf{4.15}      & \textbf{4}  \\  \hline
make sure you are on 24 $<$UNK$>$ wifi nice little printer must have unit cost too high nice $<$UNK$>$  & 
\begin{wrapfigure}{L}{0.06\textwidth} 
\vspace{-5.0mm}
\begin{minipage}{0.6\textwidth}
\includegraphics[width=0.1\linewidth,  height=0.0375\textheight]{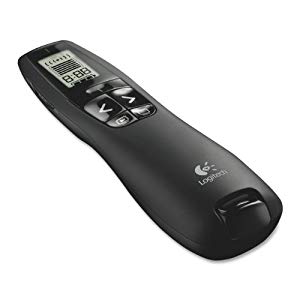}\\\\
\end{minipage} 
\end{wrapfigure}
\vspace{-1mm}
great a really nice little remote what a treat for powerpoint presentations only on some $<$UNK$>$ simple perfection (...)    &      \textbf{4.67}      & \textbf{5}  \\
\hlinewd{1pt}
\end{tabular}
}
\end{table*}

\subsubsection{Representation Alignment}

\begin{figure}[t]
\vspace{-5mm}
  \begin{subfigure}{\linewidth}
    \centering
    \includegraphics[width=7.5cm,height=1.5cm]{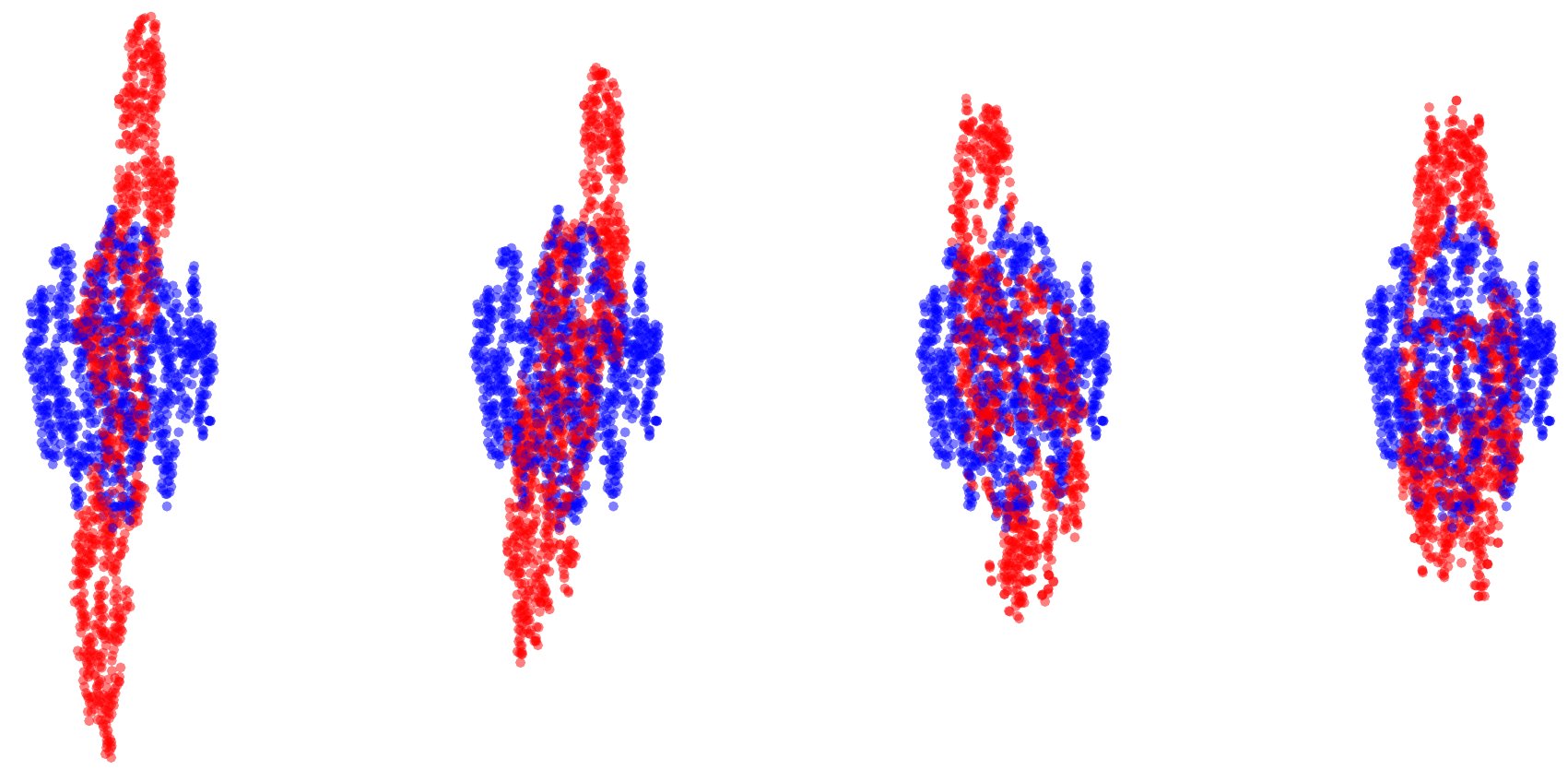}
  \end{subfigure}
  \begin{subfigure}{\linewidth}
    \centering
    \includegraphics[width=7.5cm,height=1.5cm]{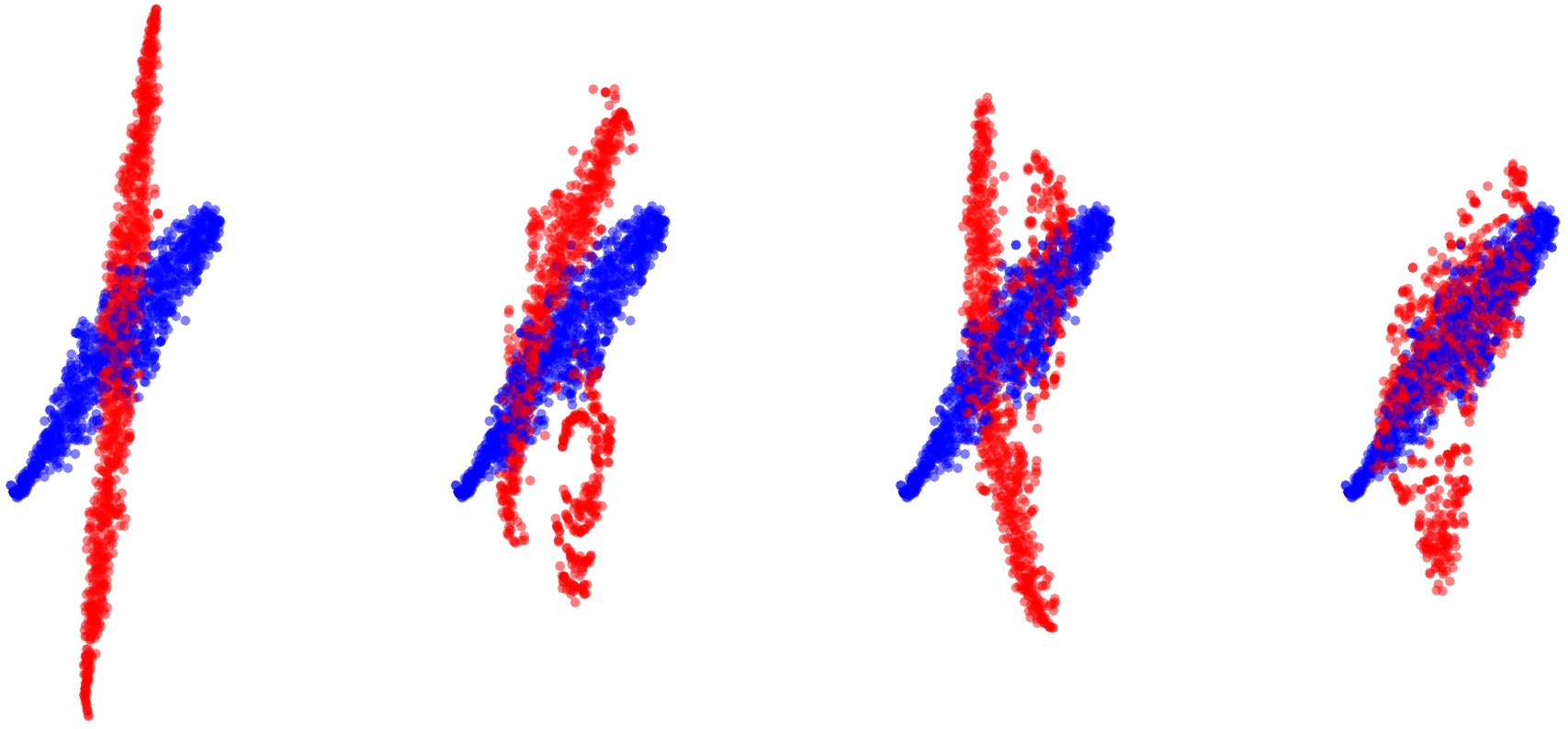}
  \end{subfigure}  
  \caption{t-SNE~\cite{maaten2008visualizing} visualizations of source (blue) and target (red) domain representations from UI-DAN (DM$\rightarrow$MI (top), HK$\rightarrow$OP (bottom)) at the 0$\textsuperscript{th}$ (no adaptation), the interaction representation adaptation at the 50$\textsuperscript{th}$, 100$\textsuperscript{th}$ and 200$\textsuperscript{th}$ epochs.}  
 \label{fig:vis_embeddings}
\end{figure}

\begin{figure}[t]
\center
    \includegraphics[width=0.7\columnwidth]{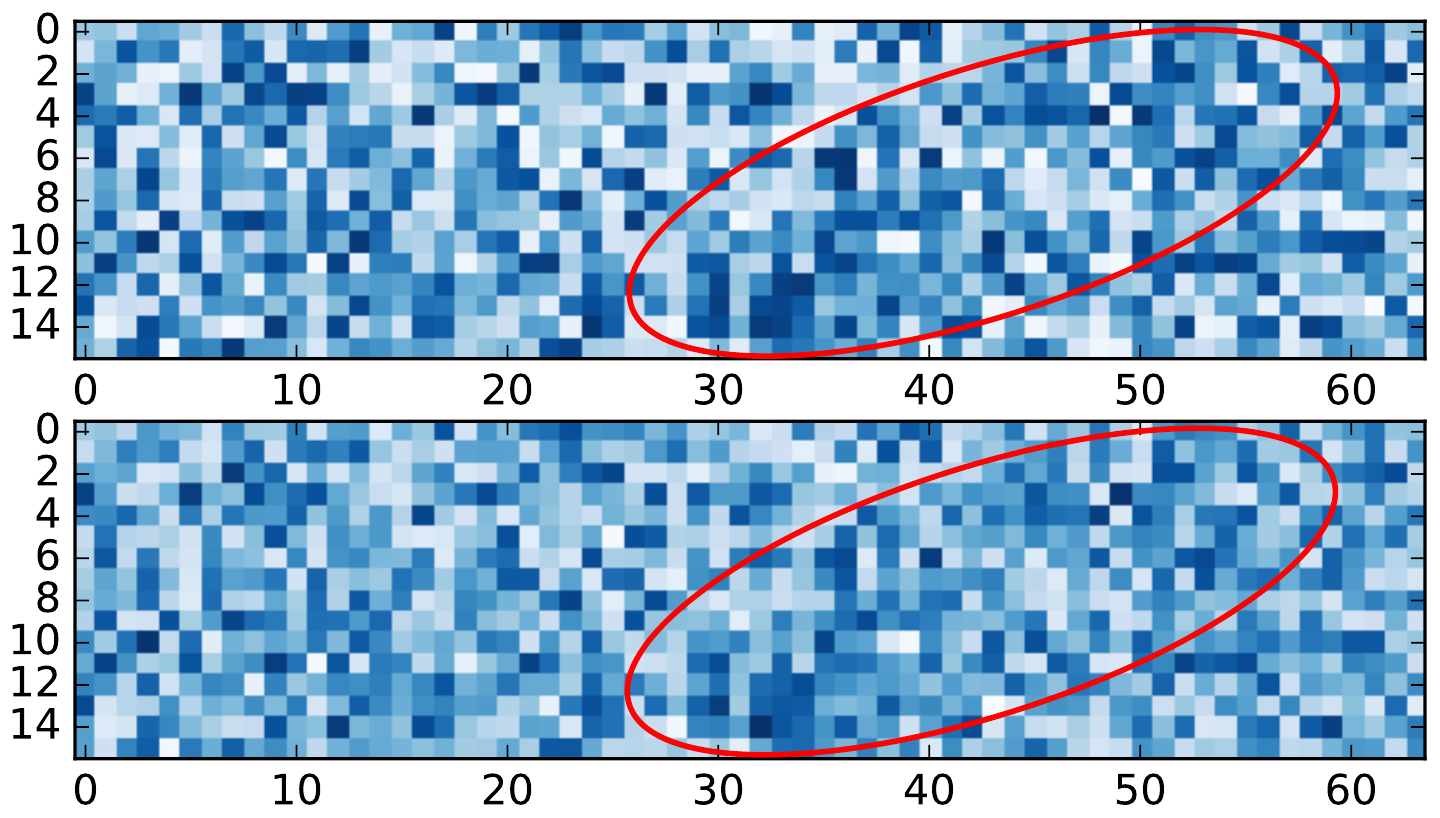}\\
  \caption{\small The visualization of weights $\Theta_f^s, \Theta_f^t \in \mathbb{R}^{512\times512}$ for source (top) and target (bottom) domains. The model is trained with DM$\rightarrow$MI domain pair, we only show the top-left 16$\times$64 of weight matrix for readability. The red circles highlight the patterns that shared by the weights of source and target domains.}  
 \label{fig:vis_weights}
 \vspace{-5pt}
\end{figure}

To examine the extent to which the adversarial objective aligns the source
and target latent representations, we randomly selected 2,000 test samples (1,000 from the source
and 1,000 from the target domain) for extracting latent
representations with $G_f$ at different epochs.
Fig.~\ref{fig:vis_embeddings} visualizes the source and target domain
representations. The source domain models' parameters are not updated during
the adversarial training of the target generators.  Comparing the
representations at the 0$\textsuperscript{th}$ epoch (no adaptation) and
50$\textsuperscript{th}$, 100$\textsuperscript{th}$,
200$\textsuperscript{th}$ epochs, we can find that the distance between
the latent representations of the source and target domains is decreasing during adversarial learning, making target representations more indistinguishable to source representations. Fig.~\ref{fig:vis_weights} shows the visualization of weights for source and target domains after training. We can observe that the weights of the target mapping function $G_f^t$ approximate those of source mapping function $G_f^s$, which again demonstrates that RecSys-DAN succeeds in aligning the representations of the source and target domains through adversarial learning.

\subsubsection{Cold-Start Recommendation}
Tab.~\ref{tab:examplel} presents some random rating prediction examples with
pre-trained RecSys-DAN models in unimodal and multimodal scenarios. We can
observe that representing users and items with reviews can effectively
alleviates the cold-start recommendation problem when ratings are completely
not available, since the proposed adversarial adaptation transfers the user,
item and their interaction representations from a labeled source domain to an
unlabeled target domain. It demonstrates the superiority of RecSys-DAN in
making preference prediction without the access to label information (i.e.,
ratings in this example). The existing recommendation
methods~\cite{lee2001algorithms, koren2008factorization, koren2010factor,
mcauley2013hidden, zheng2017joint} fail in this scenario.

\subsubsection{Running Time}
The pre-training of RecSys-DAN in the source domain took $\sim$10 epochs (avg.
69s/epoch). The adversarial training in both source and target domains took
$\sim$100 epochs to reach an equilibrium point. For inference, our model
performs as fast as baseline models, since RecSys-DAN directly adapts the
source scoring function.

\section{Conclusion}
\label{sec:con}

RecSys-DAN is a novel framework for cross-domain collaborative filtering,
particularly, the real-world cold-start recommendation problem. It learns to
adapt the user, item and user-item interaction representations from a source
domain to a target domain in an unsupervised and adversarial fashion.
Multiple generators and discriminators are designed to adversarially learn
target generators for generating domain-invariant representations. Four
RecSys-DAN instances, namely, UI-DAN, U-DAN, I-DAN, and H-DAN, are explored
by considering different scenarios characterized by the overlap of users and
items in both unimodal and multimodal settings. Experimental results
demonstrates that RecSys-DAN has a competitive performance compared to
state-of-the-art supervised methods for the rating prediction task, even with
absent preference information.

\ifCLASSOPTIONcaptionsoff
  \newpage
\fi

\begin{small}
\bibliographystyle{unsrt}
\bibliography{tnnls}
\end{small}

\end{document}